\documentclass[12pt]{article}
\usepackage{lineno}
\usepackage{graphicx}
\usepackage{amsmath}
\usepackage{hhline}
\usepackage{amssymb}
\usepackage{times}
\usepackage{cite}
\usepackage{array}
\usepackage{multirow}

\usepackage{color}
\definecolor{rltred}{rgb}{0.75,0,0}
\definecolor{rltgreen}{rgb}{0,0.5,0}
\definecolor{rltblue}{rgb}{0,0,0.75}

\newif\ifpdf
\ifx\pdfoutput\undefined
    \pdffalse          
\else
    \pdfoutput=1       
    \pdftrue
\fi
              
\ifpdf
\usepackage{thumbpdf}
\usepackage[pdftex,
        colorlinks=true,
        urlcolor=rltblue,       
        filecolor=rltgreen,     
        linkcolor=rltred,       
        pdftitle={Example File for pdflatex},
        pdfauthor={H1},
        pdfsubject={Template file},
        pdfkeywords={High-Energy Physics, Particle Physics},
        pdfpagemode=None,
        bookmarksopen=true]{hyperref}
 
\fi

\newlength{\dinwidth}
\newlength{\dinmargin}
\begin{document}

\newcommand{\pom}{{I\!\!P}}
\newcommand{\reg}{{I\!\!R}}
\newcommand{\slowpi}{\pi_{\mathit{slow}}}
\newcommand{\fiidiii}{F_2^{D(3)}}
\newcommand{\fiidiiiarg}{\fiidiii\,(\beta,\,Q^2,\,x)}
\newcommand{\n}{1.19\pm 0.06 (stat.) \pm0.07 (syst.)}
\newcommand{\nz}{1.30\pm 0.08 (stat.)^{+0.08}_{-0.14} (syst.)}
\newcommand{\fiidiiiful}{F_2^{D(4)}\,(\beta,\,Q^2,\,x,\,t)}
\newcommand{\fiipom}{\tilde F_2^D}
\newcommand{\ALPHA}{1.10\pm0.03 (stat.) \pm0.04 (syst.)}
\newcommand{\ALPHAZ}{1.15\pm0.04 (stat.)^{+0.04}_{-0.07} (syst.)}
\newcommand{\fiipomarg}{\fiipom\,(\beta,\,Q^2)}
\newcommand{\pomflux}{f_{\pom / p}}
\newcommand{\nxpom}{1.19\pm 0.06 (stat.) \pm0.07 (syst.)}
\newcommand {\gapprox}
   {\raisebox{-0.7ex}{$\stackrel {\textstyle>}{\sim}$}}
\newcommand {\lapprox}
   {\raisebox{-0.7ex}{$\stackrel {\textstyle<}{\sim}$}}
\def\gsim{\,\lower.25ex\hbox{$\scriptstyle\sim$}\kern-1.30ex%
\raise 0.55ex\hbox{$\scriptstyle >$}\,}
\def\lsim{\,\lower.25ex\hbox{$\scriptstyle\sim$}\kern-1.30ex%
\raise 0.55ex\hbox{$\scriptstyle <$}\,}
\newcommand{\pomfluxarg}{f_{\pom / p}\,(x_\pom)}
\newcommand{\dsf}{\mbox{$F_2^{D(3)}$}}
\newcommand{\dsfva}{\mbox{$F_2^{D(3)}(\beta,Q^2,x_{I\!\!P})$}}
\newcommand{\dsfvb}{\mbox{$F_2^{D(3)}(\beta,Q^2,x)$}}
\newcommand{\dsfpom}{$F_2^{I\!\!P}$}
\newcommand{\gap}{\stackrel{>}{\sim}}
\newcommand{\lap}{\stackrel{<}{\sim}}
\newcommand{\fem}{$F_2^{em}$}
\newcommand{\tsnmp}{$\tilde{\sigma}_{NC}(e^{\mp})$}
\newcommand{\tsnm}{$\tilde{\sigma}_{NC}(e^-)$}
\newcommand{\tsnp}{$\tilde{\sigma}_{NC}(e^+)$}
\newcommand{\st}{$\star$}
\newcommand{\sst}{$\star \star$}
\newcommand{\ssst}{$\star \star \star$}
\newcommand{\sssst}{$\star \star \star \star$}
\newcommand{\tw}{\theta_W}
\newcommand{\sw}{\sin{\theta_W}}
\newcommand{\cw}{\cos{\theta_W}}
\newcommand{\sww}{\sin^2{\theta_W}}
\newcommand{\cww}{\cos^2{\theta_W}}
\newcommand{\trm}{m_{\perp}}
\newcommand{\trp}{p_{\perp}}
\newcommand{\trmm}{m_{\perp}^2}
\newcommand{\trpp}{p_{\perp}^2}
\newcommand{\alp}{\alpha_s}

\newcommand{\alps}{\alpha_s}
\newcommand{\sqrts}{$\sqrt{s}$}
\newcommand{\LO}{$O(\alpha_s^0)$}
\newcommand{\Oa}{$O(\alpha_s)$}
\newcommand{\Oaa}{$O(\alpha_s^2)$}
\newcommand{\PT}{p_{\perp}}
\newcommand{\JPSI}{J/\psi}
\newcommand{\sh}{\hat{s}}
\newcommand{\uh}{\hat{u}}
\newcommand{\MP}{m_{J/\psi}}
\newcommand{\PO}{I\!\!P}
\newcommand{\xbj}{x}
\newcommand{\xpom}{x_{\PO}}
\newcommand{\ttbs}{\char'134}
\newcommand{\xpomlo}{3\times10^{-4}}  
\newcommand{\xpomup}{0.05}  
\newcommand{\dgr}{^\circ}
\newcommand{\pbarnt}{\,\mbox{{\rm pb$^{-1}$}}}
\newcommand{\gev}{\,\mbox{GeV}}
\newcommand{\WBoson}{\mbox{$W$}}
\newcommand{\fbarn}{\,\mbox{{\rm fb}}}
\newcommand{\fbarnt}{\,\mbox{{\rm fb$^{-1}$}}}
%
%
\newcommand{\qsq}{\ensuremath{Q^2} }
\newcommand{\gevsq}{\ensuremath{\mathrm{GeV}^2} }
\newcommand{\et}{\ensuremath{E_t^*} }
\newcommand{\rap}{\ensuremath{\eta^*} }
\newcommand{\gp}{\ensuremath{\gamma^*}p }
\newcommand{\dsiget}{\ensuremath{{\rm d}\sigma_{ep}/{\rm d}E_t^*} }
\newcommand{\dsigrap}{\ensuremath{{\rm d}\sigma_{ep}/{\rm d}\eta^*} }
\newcommand{\GeV}{\rm GeV}
\newcommand{\TeV}{\rm TeV}
\newcommand{\pb}{\rm pb}
\newcommand{\cm}{\rm cm}

\newcommand{\dsdE}{${\rm d}\sigma / {\rm d}E_T^{\gamma}$ }
\newcommand{\dsdeta}{${\rm d}\sigma / {\rm d}\eta^{\gamma}$ }
\newcommand{\Etg}{$E_T^{\gamma}$ }
\newcommand{\etag}{$\eta^{\gamma}$ }
%

\def\Journal#1#2#3#4{{#1} {\bf #2} (#3) #4}
\def\NCA{\em Nuovo Cimento}
\def\NIM{\em Nucl. Instrum. Methods}
\def\NIMA{{\em Nucl. Instrum. Methods} {\bf A}}
\def\NPB{{\em Nucl. Phys.}   {\bf B}}
\def\PLB{{\em Phys. Lett.}   {\bf B}}
\def\PRL{\em Phys. Rev. Lett.}
\def\PRD{{\em Phys. Rev.}    {\bf D}}
\def\ZPC{{\em Z. Phys.}      {\bf C}}
\def\EJC{{\em Eur. Phys. J.} {\bf C}}
\def\CPC{\em Comp. Phys. Commun.}

\begin{titlepage}
\begin{figure}[!t]
DESY 04--118 \hfill ISSN 0418--9833 \\
July 2004
\end{figure}

\vspace*{1cm}

\begin{center}
\begin{Large}

{\bf Measurement of Prompt Photon Cross Sections in Photoproduction at HERA 
}

\vspace{2cm}

H1 Collaboration

\end{Large}
\end{center}

\vspace{2cm}

\begin{abstract}
\noindent
   Results are presented on the photoproduction of isolated prompt photons,
   inclusively and
    associated with jets, in the $\gamma p$ center of mass energy range
 $142 < W < 266$~GeV.
  The cross sections are measured for the transverse momentum
 range of the photons
    $5 < E_T^{\gamma} < 10$~GeV
 and for associated jets with
   $E_T^{jet} > 4.5$ GeV.
 They are measured differentially as a function of
    $E_T^{\gamma}, E_T^{jet}$, the pseudorapidities \etag and $\eta^{jet}$ 
  and estimators of the momentum fractions
 $x_{\gamma}$ and  $x_{p}$
 of the incident photon and proton carried by the constituents
 participating in the hard process.
 In order to further investigate the underlying dynamics,
 the angular correlation between the prompt photon and the jet
 in the transverse plane is studied.
Predictions by perturbative QCD calculations
in next to leading order are about $30\%$ below the inclusive prompt
photon data after corrections for hadronisation and multiple interactions,
 but are in reasonable agreement with the results for prompt photons
 associated with jets.
 Comparisons with the predictions of
 the event generators PYTHIA and
  HERWIG are also presented.
\end{abstract}

\vspace{1.5cm}

\begin{center}
To be submitted to Eur. Phys. J. C 
\end{center}

\end{titlepage}
%
%

\begin{flushleft}

A.~Aktas$^{10}$,               
V.~Andreev$^{26}$,             
T.~Anthonis$^{4}$,             
A.~Asmone$^{33}$,              
A.~Babaev$^{25}$,              
S.~Backovic$^{37}$,            
J.~B\"ahr$^{37}$,              
P.~Baranov$^{26}$,             
E.~Barrelet$^{30}$,            
W.~Bartel$^{10}$
% ...

\bigskip{\it
% ...
$ ^{4}$ Inter-University Institute for High Energies ULB-VUB, Brussels;
          Universiteit Antwerpen, Antwerpen; Belgium$^{ c}$ \\
% ...
$ ^{10}$ DESY, Hamburg, Germany \\
% ...
$ ^{25}$ Institute for Theoretical and Experimental Physics,
          Moscow, Russia$^{ l}$ \\
 $ ^{26}$ Lebedev Physical Institute, Moscow, Russia$^{ e}$ \\
% ...
$ ^{30}$ LPNHE, Universit\'{e}s Paris VI and VII, IN2P3-CNRS,
          Paris, France \\
% ...
$ ^{33}$ Dipartimento di Fisica Universit\`a di Roma Tre
          and INFN Roma~3, Roma, Italy \\
% ...
$ ^{37}$ DESY, Zeuthen, Germany \\
% ...
$ ^c$ Supported by FNRS-FWO-Vlaanderen, IISN-IIKW and IWT
      and  by Interuniversity Attraction Poles Programme,
      Belgian Science Policy \\
% ...
$ ^e$ Supported by the Deutsche Forschungsgemeinschaft \\
% ...
$ ^l$ Partially Supported by Russian Foundation
      for Basic Research, grant    no. 00-15-96584 \\
}

\end{flushleft}

\newpage

\section{Introduction}
\noindent

\begin{figure}[ht] \unitlength 1cm
  \begin{center}
\begin{picture}(16,10)
\put(1.,1.){\includegraphics[width=0.9\textwidth]{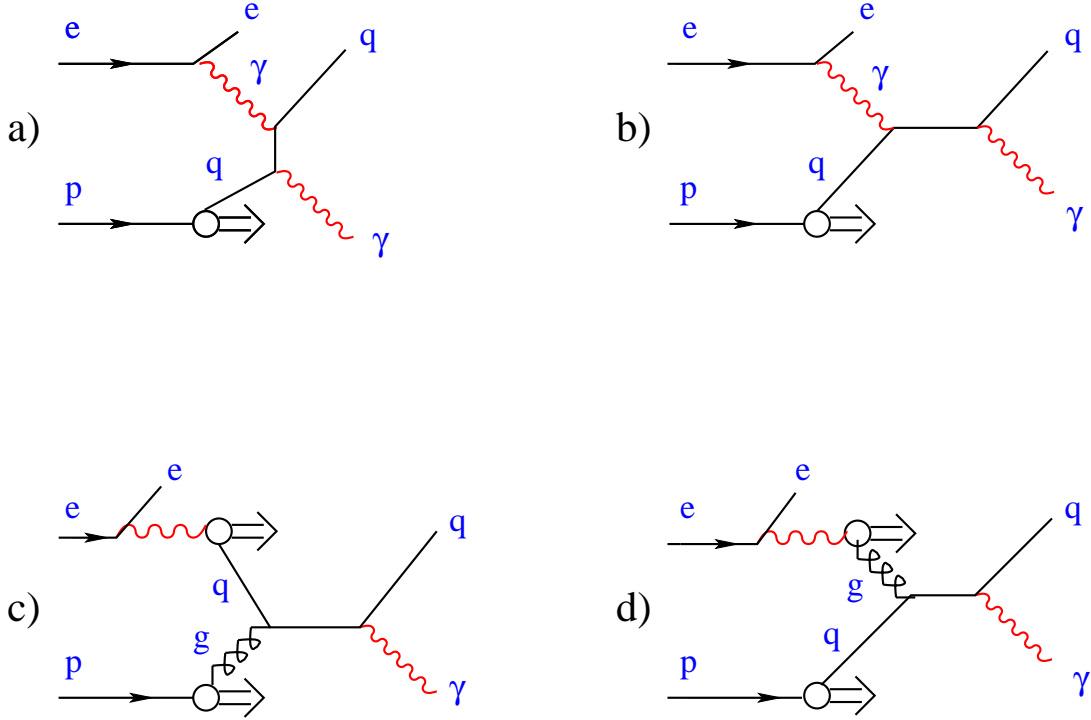}}
 \end{picture}
  \caption{
 Examples of leading order diagrams producing prompt photons. 
 Direct photon interactions a), b) and resolved photon interactions c),d).  
 }
  \label{feyn}
 \end{center}
\end{figure}

Prompt photon emission in hadronic interactions is a sensitive probe
of QCD dynamics and partonic structure, providing complementary
information to the study of jet production. Although cross sections are
smaller in the prompt photon case, an isolated photon at large transverse
energy \Etg can be related directly to the partonic event structure. This
contrasts with jet measurements, where the partonic structure is obscured
by the non-perturbative hadronisation process. Furthermore, experimental
uncertainties associated with the transverse energy measurement
are smaller for the electromagnetic showers initiated by photons than
for the measurement of jets of hadrons.
This paper presents results from the H1 experiment at HERA
on the photoproduction of isolated prompt photons, both inclusively and
in association with a jet.
Here photons are called ``prompt'' if they are
coupled to the interacting quarks (see Fig.~\ref{feyn}), in
contrast to photons which are produced as decay products of hadronic
particles.

Next to leading order (NLO) perturbative QCD (pQCD) calculations have proved
broadly successful in reproducing measured jet production rates from various
colliders, provided hadronisation corrections are applied. In
contrast, discrepancies have been observed between data on prompt photon
production and NLO pQCD calculations
in $pp$, $\bar{p}p$, $pN$ (see e.g.~\cite{Apanasevich:1998ki})   and in
$\gamma p$ \cite{Breitweg:1999su}
interactions.
These discrepancies can be
reduced by introducing intrinsic transverse momentum $k_T$
to the incoming partons of the proton
~\cite{Apanasevich:1998ki,Chekanov:2001aq}
or by soft gluon resummations~\cite{Lai:1998xq}.
NLO pQCD calculations of prompt photon production in
photoproduction~\cite{Fontannaz:2001ek,Fontannaz:2001nq,Krawczyk:2001tz,Zembrzuski:2003nu,Gordon:1994km}
are available, in which the incident photon interacts with
the partons of the proton either directly (Fig.~\ref{feyn}a,b)
or via its ``resolved'' partonic structure (Fig.~\ref{feyn}c,d).
This paper investigates the extent to which fixed order
NLO calculations are able to describe the new data.
The data are also compared with the predictions of
the event generators PYTHIA~\cite{Sjostrand:2000wi}
and HERWIG~\cite{Corcella:1999qn} and with data 
on inclusive prompt photon production
from the ZEUS collaboration~\cite{Breitweg:1999su}.

\section{Strategy of Prompt Photon Measurement}

The photoproduction process is initiated
 by quasi-real photons, which  are produced in small angle $ep$ scattering,
  where the scattered electron\footnote{The term
  ``electron'' is used for both electrons and positrons.}
 escapes into the beam pipe.
  The requirement of
  a significant energy loss in the electron
  beam direction, together with the condition that no electrons
  are found in the selected events, suppresses contributions
  from  neutral current (NC)
  deep inelastic scattering (DIS). 

  Photons are
 identified  in the H1 liquid argon (LAr) calorimeter~\cite{Andrieu:1993kh}
  as compact electromagnetic clusters
  with no track associated to them.
  The main experimental difficulty is the separation of the prompt photons 
  from hadronic backgrounds, in particular 
  $\pi^0$~meson decays, since at high energy the
  decay photons cannot be reconstructed separately in the calorimeter.
  Since the $\pi^0$~mesons are predominantly produced in jets,
  this background is strongly reduced by requiring the photon candidates
  to be isolated from other particles.

  After the selection cuts described below, the $\pi^0$ background
 is still of similar size to
  the prompt photon signal.
  To extract the signal, different shower shape variables are
      combined to form a discriminator function which is fitted with 
  a sum of contributions from simulated photons, $\pi^0$ and $\eta$~mesons.
 The fit is done double-differentially in bins of 
 transverse energy \Etg and pseudorapidity $\eta^{\gamma}$.
  \footnote{The pseudorapidity $\eta$ of an object with polar angle $\theta$
   is given by
   $\eta = -\ln \; \tan (\theta/2)$, where $\theta$ 
   is measured with respect to the $z$ axis
   given by the proton beam direction.
   Transverse energies are also measured with respect to this axis  
   unless otherwise specified.}

The number of selected 
 prompt photon events is corrected
for detector effects
  by detailed simulations
  of prompt photon production in the H1 detector, using the
  event generators PYTHIA and HERWIG.

\section{Event Selection}

The data were collected in the years 1996-2000 with
 the H1 detector~\cite{Abt:hi}
 at HERA in data taking periods where electrons or positrons with 
 energy $E_e = 27.6$ GeV collided with protons of energies
 $E_p = 820$ GeV or $E_p = 920$ GeV.
The data correspond 
to an integrated luminosity of $105~\pb^{-1}$ of which
 $28.8~\pb^{-1}$ and $61.3~\pb^{-1}$ were recorded in $e^+p$ interactions  
 at center of mass energies $\sqrt{s} = 301~\GeV$ and
 $\sqrt{s} = 319~\GeV$, respectively,
 and $14.9~\pb^{-1}$ were recorded 
 in $e^-p$ interactions at $\sqrt{s} = 319~\GeV$.

The events are triggered by compact energy depositions
 in the LAr calorimeter,
consistent with electron or photon showers.
The trigger efficiency is $\approx 60\%$
 at photon energies of 5 GeV, reaching 
$100\%$ at $\approx 12$~GeV.

The main requirements for the event selection 
are the following~\cite{rachid}.
\begin{itemize}
\item 
A compact electromagnetic energy cluster,
consistent with a $\gamma$ shower,
 is reconstructed in the LAr 
calorimeter in the range  $-1 < \eta^{\gamma} < 0.9$ and
 \Etg $> 5$~GeV.
For the data at $\sqrt{s} = 301~\GeV$ a threshold \Etg $> 7$~GeV
is required.\footnote{The data at $\sqrt{s} = 301~\GeV$ were taken prior
to the upgrade of
 the LAr electronics, after which lower trigger thresholds became possible.}
The $\eta^{\gamma}$ range corresponds to the central barrel
region of the LAr calorimeter~\cite{Andrieu:1993kh}.

\item
No track is allowed to point to this cluster within a distance of 25 cm
in the plane transverse to the track
 at the calorimeter surface. 
 
\item
Events with electron candidates
 in the LAr calorimeter
 or in the backward calorimeter SPACAL~\cite{Nicholls:1995di} 
 are rejected.
 This restricts the virtuality of the exchanged photon
 to $Q^2 < 1$ GeV$^2$ and suppresses contributions from
 radiative DIS and  QED Compton
 processes.

\item
At least two tracks are required in
 the central tracker~\cite{Abt:hi,Burger:eb},
which covers the angular range $20^{\circ} < \theta < 160^{\circ}$.
This cut assures
good reconstruction of the event vertex, which is required to be 
 within $\pm 35$ cm in $z$ of the nominal vertex position. 

\item
 The inelasticity $y = W^2/s$, 
 where $W$ is the  $\gamma p$ center of mass energy,
 is evaluated as $y = \sum{(E - p_z)}/2E_e$.   
 Here the sum runs over all detected final state particles.
  The required range  $0.2 < y < 0.7$ corresponds to 
 $142 < W < 266$~GeV
 for
 $E_p = 920$~GeV.
 The cut at high $y$ reduces NC DIS background. The cut at low $y$
removes beam gas background.

\item
 The $\gamma$ candidate is required to be isolated.
 The transverse energy, $E_T^{cone}$, in a cone
 around the $\gamma$ candidate, given by 
 distances below 1 unit in the  $(\eta - \phi)$ plane,
 is required to be less than
 10\% of \Etg.
 A further cut to remove non-prompt photon background is based 
 on the shower shape, as described in section~\ref{sec:signal}.

\item
 Associated jets are reconstructed using the inclusive $k_T$
 algorithm~\cite{Ellis:tq} with a distance parameter $D=1$
 in the $(\eta-\phi)$ plane.
 The jets are selected in the 
jet energy and pseudorapidity ranges
$E_T^{jet} > 4.5$ GeV
 and $-1 < \eta^{jet} < 2.3$, respectively.
 If more than one jet is found, only the jet with the highest
$E_T^{jet}$ is considered. 
         
\end{itemize}
  
\section{Signal Extraction}
\label{sec:signal}

 The prompt photon signal is extracted
 using a shower shape analysis based on the expectation
 that showers initiated by photons
 are typically narrower than
$\pi^0$ or $\eta$ initiated showers,
  with less energy deposited in the first
  calorimetric layer on average.
 This procedure  exploits the fine segmentation of the electromagnetic
 section  of the LAr calorimeter, which has
transverse cell sizes varying between  
 about $5 \times 7$ cm$^2$ and $7 \times 13$ cm$^2$
 in the central barrel region.
 This part of the calorimeter has three layers in depth,
 corresponding to $\approx 20$ radiation lengths.
 The first layer
 has a thickness of about 3 radiation lengths. 

 Three observables are used to discriminate against
 background.
 The mean transverse shower radius is given by
      $R = \sum_{i} r_i \varepsilon_i / \sum_{i} \varepsilon_i$,
  where $r_i$ is the transverse distance of cell $i$ with energy
  density
 $\varepsilon_i$ measured 
   with respect to the axis from the event vertex to the center of gravity
   of the  $\gamma$ candidate cluster.
  The shower hot core fraction ($HCF$) is the
 largest energy fraction of the candidate shower 
   which is contained in 4 or 8 contiguous cells
 (depending on the calorimeter granularity)
 including the cell of highest energy.
 Finally the first layer fraction ($FLF$) is the energy fraction
 of the shower contained in the first,
 i.e. closest to the beam,  layer of cells of the calorimeter. 
 The observables $R$ and $FLF$ are expected to be smaller
and $HCF$ to be larger
  for  the prompt photon showers than for background
  showers.  
 
\begin{figure}[ht] \unitlength 1cm
  \begin{center}
\begin{picture}(16,20)
\put(0.,9.5){\includegraphics[width=0.45\textwidth]{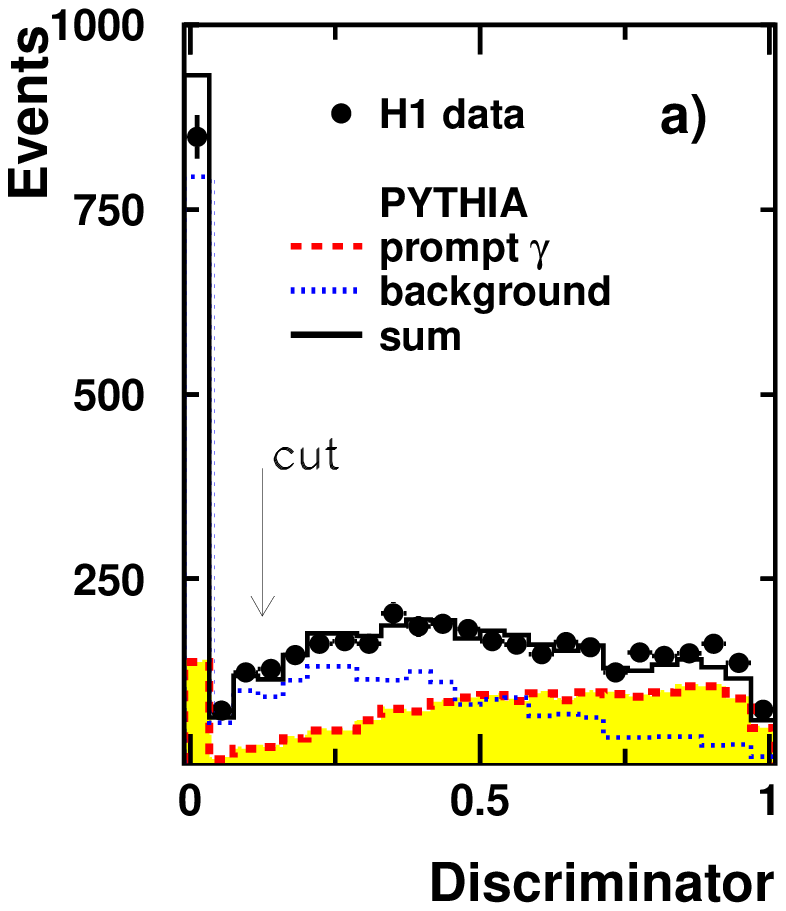}}
\put(8.,9.5){\includegraphics[width=0.45\textwidth]{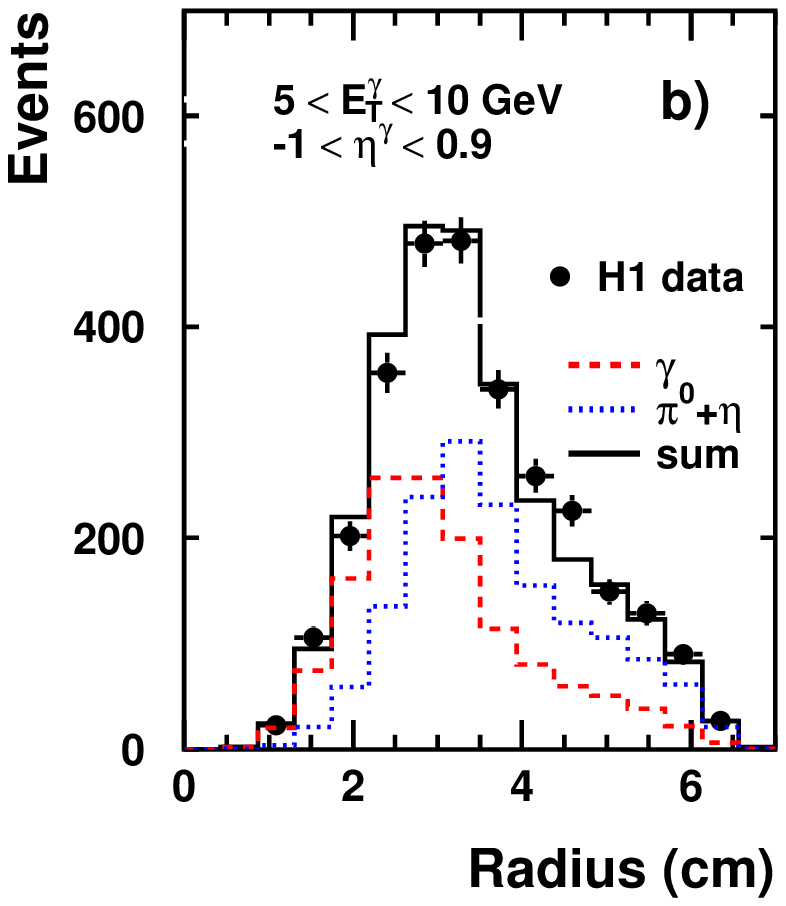}}
\put(0.,0.5){\includegraphics[width=0.45\textwidth]{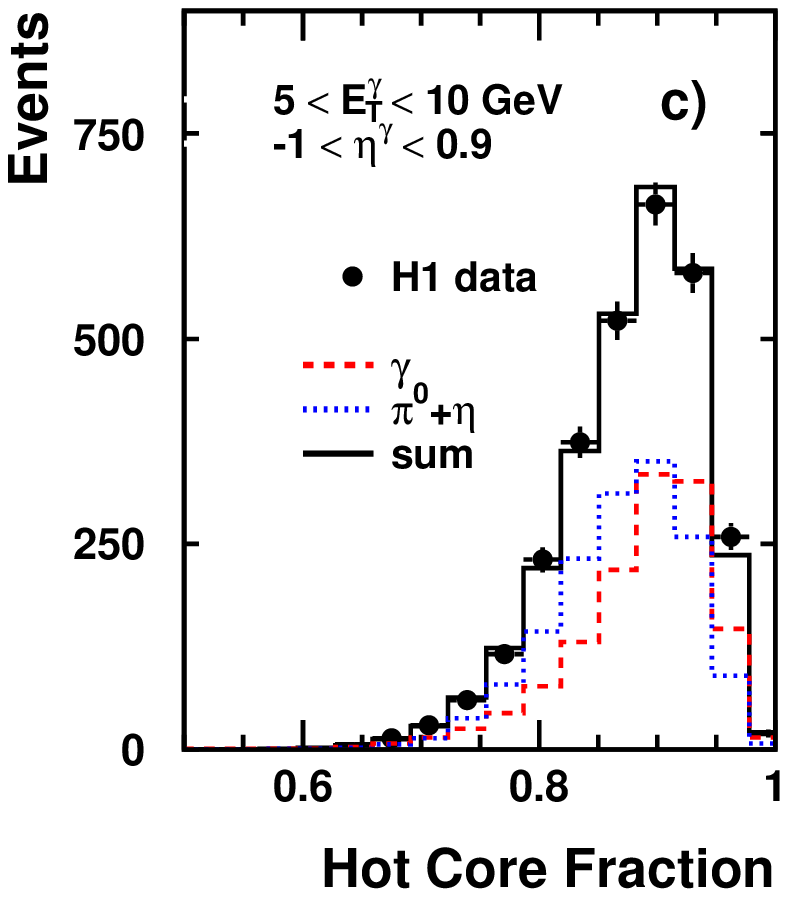}}
\put(8.,0.5){\includegraphics[width=0.45\textwidth]{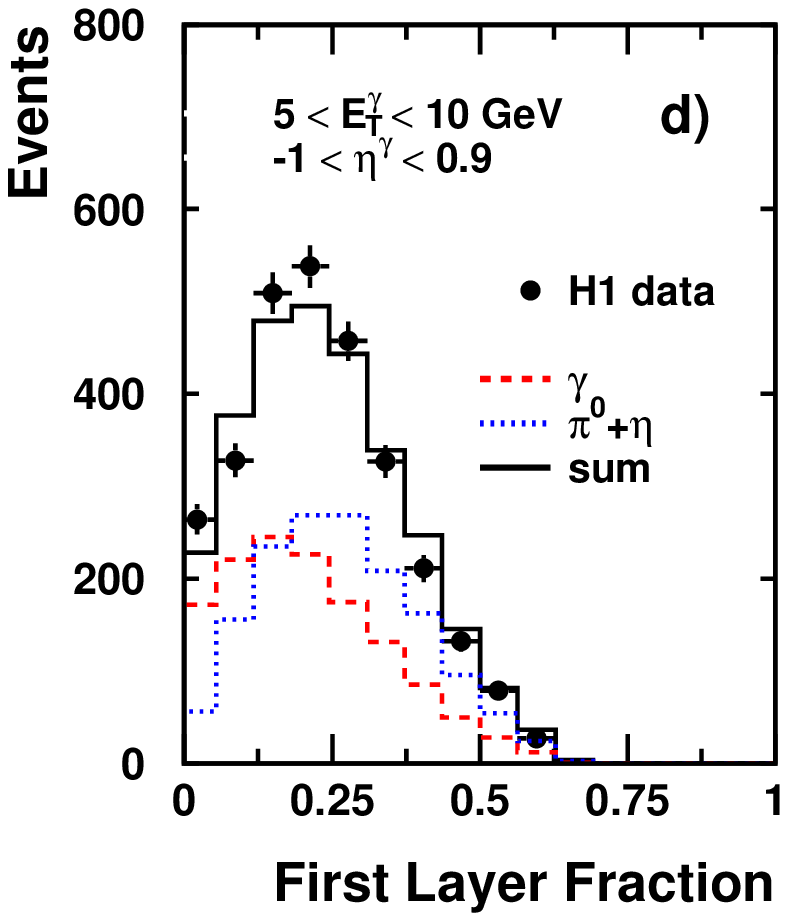}}
 \end{picture}
  \caption{
  a) Distribution in the discriminator $d$, with the analysis cut indicated,
 for data (solid points),
     and PYTHIA normalised to the data (solid line), with
     prompt photons (dashed) and the sum of the background
   contributions (dotted).
  b) to d) Distributions of the mean transverse shower radius $R$, 
   the hot core fraction $HCF$ and the first layer fraction $FLF$
 for all selected photon candidates
   (data points).
The contributions determined for photons (dashed lines),
   background ($\pi^0 + \eta$, dotted lines) and the sum (solid lines)
 are also shown.
 }
  \label{fig:signal}
 \end{center}
\end{figure}

 To discriminate between photon and background showers,
 probability densities
 for the three observables $R$, $HCF$ and $FLF$
 are determined by simulation of photons and
 $\pi^0$~mesons~\cite{rachid}.
 The products of these three densities are used as likelihood functions.
For each measured event a discriminator ($d$) is formed
by the likelihood for photons divided by the sum of the 
likelihoods for photons and $\pi^0$~mesons.
The discriminator $d$ produces
larger values for prompt photons 
 than for the pairs of photons from 
$\pi^0$ or $\eta$~meson decays.
 This can be seen in Fig.~\ref{fig:signal}a,
 where the measured distribution of $d$ is shown together with 
the prompt photon and background components estimated from the PYTHIA 
Monte Carlo simulation.
 The lowest bin in Fig.~\ref{fig:signal}a contains mostly background
 ($\approx 90\%$) which 
 is composed of
 65\% $\pi^0$,  30\% $\eta$ and 5\% other particles. 
 After a cut $d > 0.125$, 
 the remaining background  ($\approx 50\%$)
 is composed of
 94\% $\pi^0$,  5\% $\eta$ and about 1\% other particles.
   Fig.~\ref{fig:signal}a shows that
  the distribution in the discriminator $d$
 of signal plus background is well predicted by PYTHIA.

The contribution of prompt photons is determined by maximum
likelihood fits of simulated photon, $\pi^0$ and $\eta$~meson
discriminator distributions to the data distribution for
$d > 0.125$. Each measurement presented in section 6 is obtained
by summing the results of such fits performed independently
in $6 \times 6$ bins of $\eta^{\gamma}$ and $E_T^{\gamma}$.
In this procedure, only the $\eta/\pi^0$ ratio (on average 5\%
after the selection requirements) is taken from PYTHIA.

  The measured distributions of $R$, $HCF$ and  $FLF$
 for the full $\eta^{\gamma}$ and \Etg range
 are shown in Fig.~\ref{fig:signal}b-d
  together with the contributions of
  photons and $\pi^0$ and $\eta$ background,
  the normalisations of which are taken from the fits to the discriminator 
distributions described above.
 The data distribution is well
   described by the extracted signal and background components.
   The discrimination between signal and background becomes weaker at
   high \Etg, where the $R$ and $HCF$ distributions
   of $\pi^0$~mesons and photons become more similar to each other.
    Therefore events with $E_T^{\gamma} > 10$~GeV are not included
  in the results presented below.

\section{Monte Carlo Generators and Corrections to the Data}
\label{sec:corr}

 The event generators
 PYTHIA~\cite{Sjostrand:2000wi}
  and HERWIG~\cite{Corcella:1999qn} are used
\footnote{PYTHIA 6.15/70 and HERWIG 6.1 are used
 with default multiple interactions
 (MSTP(82)=1 for PYTHIA, PRSOF=1, BTCLM=1 for HERWIG).  
}
 to correct the observed event yields
  for apparatus effects by means of a full simulation 
  of the H1 detector.
 The average corrections
 from the two generators are applied.
  Both generators are based on leading order (LO) QCD matrix elements
  and leading log parton showers.
Hadronisation is provided in PYTHIA by
 Lund string fragmentation~\cite{Sjostrand:1993yb} and in HERWIG by
 the decay of colourless parton clusters.
 Both generators model additional soft remnant-remnant interactions,
 termed multiple parton interactions (m.i.) in the following.  
    The
     GRV(LO)~\cite{Gluck:1991jc}  parton densities are used 
    for the photon and the proton.
    In contrast to HERWIG, PYTHIA simulates  
radiation of photons from the electron line
 and photon production via fragmentation
of final state quarks and gluons in di-jet events.
In PYTHIA the parameter describing the 
intrinsic $k_T$ of initial state
    partons in the proton is $k_0$ = 1 GeV (default)
 leading to $<k_T> = 0.9$ GeV. HERWIG predictions are shown for the
 default value  $k_T = 0$. 

A correction factor of about 1.04 is applied to the
 $\sqrt{s} = 301$~GeV data
 in order to combine with the data at 
 $\sqrt{s} = 319$~GeV. The final results are presented for
  $\sqrt{s} = 319$~GeV.
   Background from DIS events where the scattered electron
   fakes the prompt photon signatures due to
  tracker inefficiency
   leads to a subtraction of 3\% 
    in the lowest \etag bin at
   high \Etg , and is negligible otherwise.  
  Following the selection criteria described in section 3, other
  sources of background are negligible.

\section{Systematic Uncertainties}

 The following sources of systematic errors are considered.

\begin{itemize}

\item
   The dominant systematic errors are due to possible imperfections in the
  simulation of the shower shapes.
The quality of the simulated distributions of $R$, $HCF$ and $FLF$
  is tested by comparing simulations of
  electrons with electron candidates
 from NC DIS events.
 Differences between the data and the simulation in these distributions
 result in errors on the prompt photon cross sections ranging from
 $\pm 10\%$  to $\pm 20\%$.

\item 
 The uncertainties on the calorimeter electromagnetic and hadronic
 energy scales ($0.7\%$ to $1.5\%$ and $\approx3\%$ respectively) 
 contribute errors of about 5\% for the inclusive cross sections.
  For the case with associated jets,
 the hadronic energy uncertainty gives rise to uncertainties of 
  about 10\%.

\item
 The trigger efficiency is determined using independent triggers
 with an uncertainty of 3\%.

\item
The model dependence,
of the corrections for detector effects
 is quantified as
 half the difference ($< 6\%$ in most cases)
 between the correction factors
 obtained with PYTHIA and HERWIG.  
 Within these uncertainties the results are insensitive to reasonable
 variations of the \Etg dependence and of the underlying event activity
in PYTHIA.  
\end{itemize}

   An  overall normalisation
   uncertainty of $\pm 1.5\%$ on the luminosity measurement is not
   included in the results.
 The total systematic errors are obtained 
 by adding the different systematic errors in quadrature.
For further details see~\cite{rachid}.
     
\section{Results}
  The results are presented in tables 1 to 3 and figures~\ref{fig:pythia}
  to ~\ref{pperp}
   as bin averaged $ep$ cross sections
   in the kinematic region defined by  
$$\sqrt{s} = 319\; {\rm GeV}, \; \; 0.2 < y < 0.7, \;
 \; Q^2 < 1 \; {\rm GeV}^2 \;$$
$$  {\rm and}
 \;\;\; 5 < E_T^{\gamma} < 10 \; {\rm GeV}, \;\; -1 < \eta^{\gamma} < 0.9,\;\; 
  E_T^{cone} < \; 0.1 \cdot E_T^{\gamma}\;.$$
The inner error bars on the data points in the figures 
 indicate the statistical errors as obtained from 
the shower discriminating fits. The full error bars
  also contain
 the systematic errors added in quadrature. 

\subsection{Inclusive prompt photons}

\begin{figure}[ht] \unitlength 1cm 
\begin{center}
\begin{picture}(16,8)
\put(0.5,0.){\includegraphics[width=0.9\textwidth]{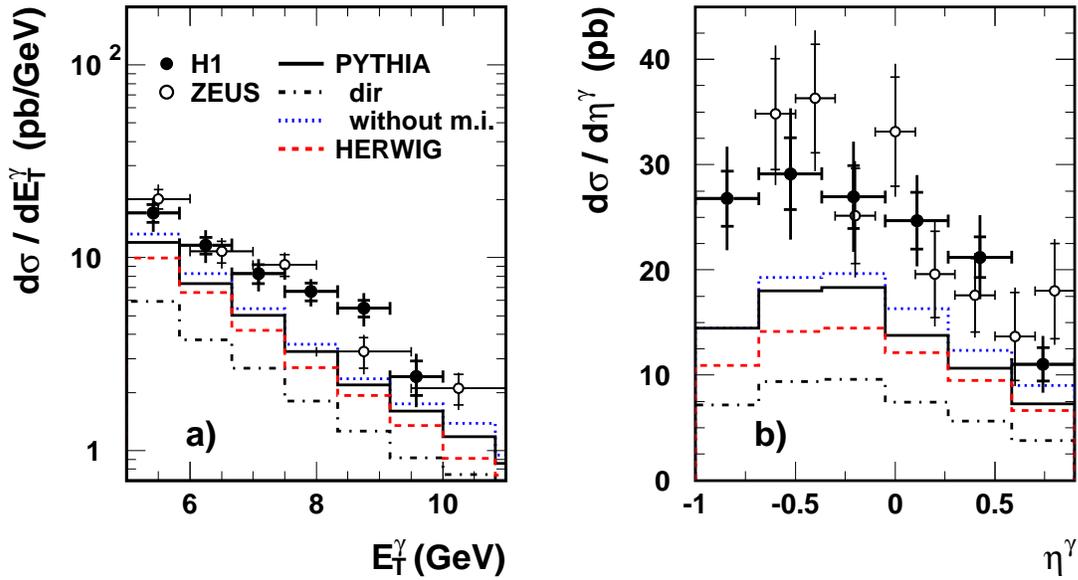}}
 \end{picture}
  \end{center}
  \caption{Inclusive prompt photon differential cross sections
\dsdE for  $-1 < \eta^{\gamma} < 0.9$ (a)
and \dsdeta for $5 <$ \Etg $< 10$ GeV (b)
 with $\sqrt{s} = 319~\GeV$ and $0.2 < y < 0.7$
 compared with the predictions of HERWIG (dashed line) and
 PYTHIA including multiple interactions
 (full line).
 The contribution of direct interactions is shown separately
 (dashed-dotted line). The full
  PYTHIA prediction without multiple interactions (dotted line)
 is also shown.
 ZEUS data~\cite{Breitweg:1999su} are shown 
 adjusted to correspond to
 $\sqrt{s} = 319~\GeV$, $0.2 < y < 0.7$ and $-1 < \eta^{\gamma} < 0.9$.
}
  \label{fig:pythia}
\end{figure}

Differential cross sections \dsdE and \dsdeta for inclusive prompt photon
 production
   are shown in Fig.~\ref{fig:pythia} and  are compared with
   the predictions of the
 PYTHIA~\cite{Sjostrand:2000wi}
  and HERWIG~\cite{Corcella:1999qn} event generators.
 The cross sections are reasonably described in shape, but 
 the predictions by PYTHIA (HERWIG)
  are low by about 40\% (50\%) in normalisation.
  Photons from fragmentation in di-jet events are only treated in the PYTHIA
  calculation,
  which explains the difference from HERWIG.
 The figure shows the full PYTHIA prediction  
  and separately the contribution of 
  direct photon interactions only.
   The PYTHIA
  prediction is also shown without multiple interactions.  
  The predictions 
  at $0 <$ \etag $< 0.9$ are about 25\% higher
 without multiple interactions,
 showing that
  the cross section is reduced by the soft underlying event activity,
 as expected~\cite{Fontannaz:2001ek} 
  due to the isolation cone condition 
   $E_T^{cone} = 0.1  \cdot E_T^{\gamma}$.
 Fig.~\ref{fig:pythia} also shows a comparison  with
 the results of the ZEUS collaboration~\cite{Breitweg:1999su}.\footnote{The ZEUS data are obtained in the somewhat different kinematic region
   $0.2 < y < 0.9$, $-0.7 < \eta^{\gamma} < 0.9$, 
   $\sqrt{s} = 301~\GeV$ and are adjusted to correspond to the H1 conditions
   using the NLO calculation~\cite{Fontannaz:2001ek}.}
The two measurements are consistent. 
     
\begin{figure}[ht] \unitlength 1cm 
  \begin{center}
\begin{center}
\begin{picture}(16,18)
\put(0.5,9.0){\includegraphics[width=0.85\textwidth]{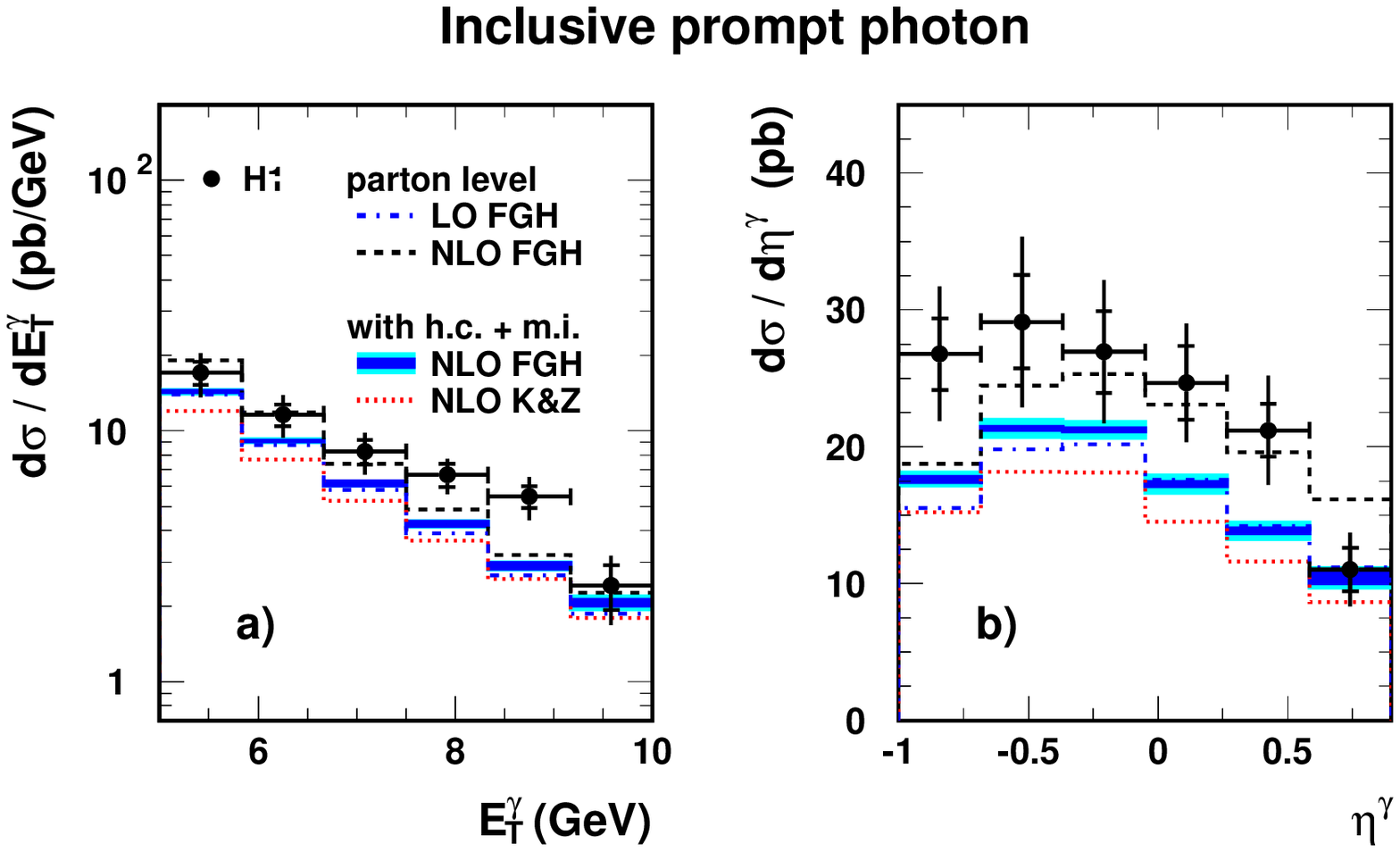}}
\put(0.5,0.){\includegraphics[width=0.85\textwidth]{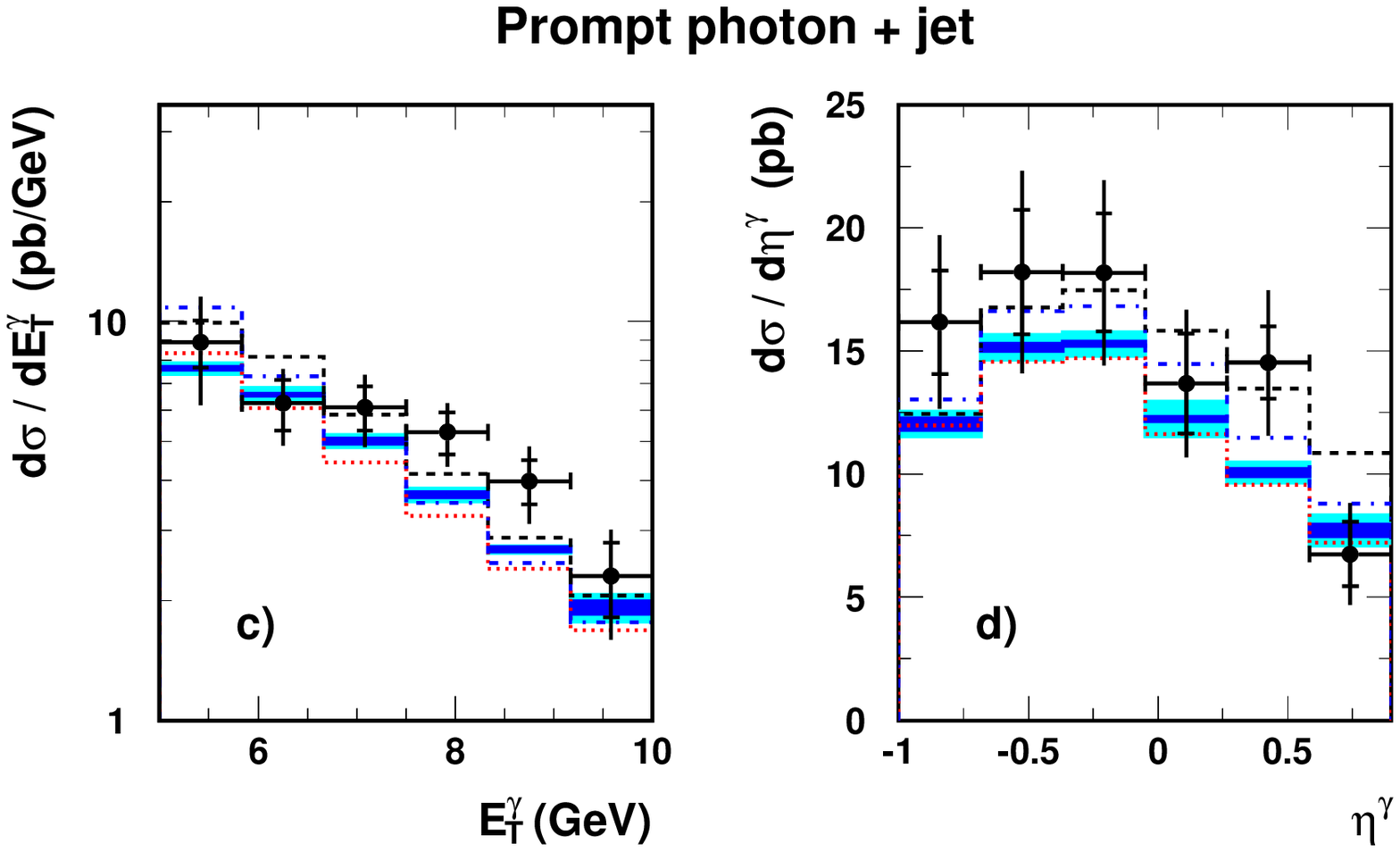}}
 \end{picture}
  \end{center}
  \caption{Inclusive prompt photon cross sections (a,b)
 for  $-1 < \eta^{\gamma} < 0.9$ 
and
 $5 <$ \Etg $< 10$~GeV
 with $\sqrt{s} = 319~\GeV$ and $0.2 < y < 0.7$,
and with an additional jet requirement      
  ($E_T^{jet} > 4.5$~GeV, $-1 < \eta^{jet} < 2.3$) (c,d).
The data are compared with NLO pQCD calculations  
(K\&Z~\cite{Zembrzuski:2003nu}, dotted line, and
FGH~\cite{Fontannaz:2001ek}, solid line).
The NLO results are    
 corrected for hadronisation and 
 multiple interaction (h.c. + m.i.) effects (see text).
 The outer error bands show
 the estimated uncertainties on these corrections
 for the example of FGH,
  added linearly
 to the
 effect of a variation of the renormalisation and factorisation scales
 in the NLO calculation from
 $0.5 \cdot E_T^{\gamma}$ to $2 \cdot E_T^{\gamma}$ (inner band).
 The FGH results are  also shown
 without corrections for h.c. and m.i. at NLO (dashed) and LO (dashed-dotted).
}
  \label{fig:res}
  \end{center}
\end{figure}

 The results are further compared in Figs.~\ref{fig:res}a,b with 
    the NLO pQCD calculations by Fontannaz, Guillet
 and Heinrich (FGH)~\cite{Fontannaz:2001ek}
and Krawczyk and Zembrzuski (K\&Z)~\cite{Krawczyk:2001tz,Zembrzuski:2003nu}.
 The two calculations are similar,
 the main difference being that only FGH apply higher order corrections
 to the resolved photon processes (Fig.~\ref{feyn}c,d).
   The final state photon can be emitted in a hard partonic process
   or can be produced in the fragmentation of a quark or gluon.
  The photon isolation requirement in the cross section definition
  suppresses the latter contribution considerably. According  to the
   calculations, the contribution to the cross sections from fragmentation
  is typically 10\%.
 Both calculations use
 the photon and proton parton density functions
    AFG~\cite{Aurenche:1994in} and
  MRST2~\cite{Martin:1999ww}, respectively,
 and 
  BFG~\cite{Bourhis:1997yu} fragmentation functions.
   The transverse energy \Etg is used for the renormalisation
   and factorisation scales.
 In order to obtain a realistic comparison of data and theory, 
 the NLO calculations are shown after correction 
 from the parton to the hadron level including the effects of
 multiple interactions (m.i.).
 These corrections are obtained using the average of PYTHIA and HERWIG,
 taking half the difference as an uncertainty estimate.
 The resulting uncertainty is 
 typically 3\% as shown by the outer error bands in Figs.~\ref{fig:res}a,b.  
 
 The FGH (K\&Z) NLO calculations
 are typically 30\% (40\%) below the data
 in most of the \Etg and \etag ranges presented in Figs.~\ref{fig:res}a,b,
 if the corrections
 for hadronisation and m.i. are applied. 
 The comparison with the parton level result of FGH
 shows that the correction factors are largest at high $\eta^{\gamma}$,
 where resolved photon interactions contribute
 most. Only here ($\eta^{\gamma} > 0.6$) the corrections improve the
 agreement with the data. 
 The NLO corrections are substantial,
  with the  NLO/LO ratio increasing from 1.2 to 1.4 for FGH
  with increasing $\eta^{\gamma}$.
 The effect on the NLO calculations of scale variations is rather small
 as 
 shown by the inner error bands in 
 Fig.~\ref{fig:res}a,b, which indicates the effect of the 
 variation of the renormalisation
 and factorisation scales from  
  $0.5 \cdot E_T^{\gamma}$ to $2 \cdot E_T^{\gamma}$.
 The predictions are
 about 10\% larger on average if
 the GRV parton density and fragmentation  
 functions~\cite{Gluck:1991jc,Gluck:1991ee} are used.   
    
\subsection{Prompt photons with jets}

\begin{figure}[ht] \unitlength 1cm
\begin{center}
\begin{picture}(16,18)
\put(0.5,0.5){\includegraphics[width=0.85\textwidth]{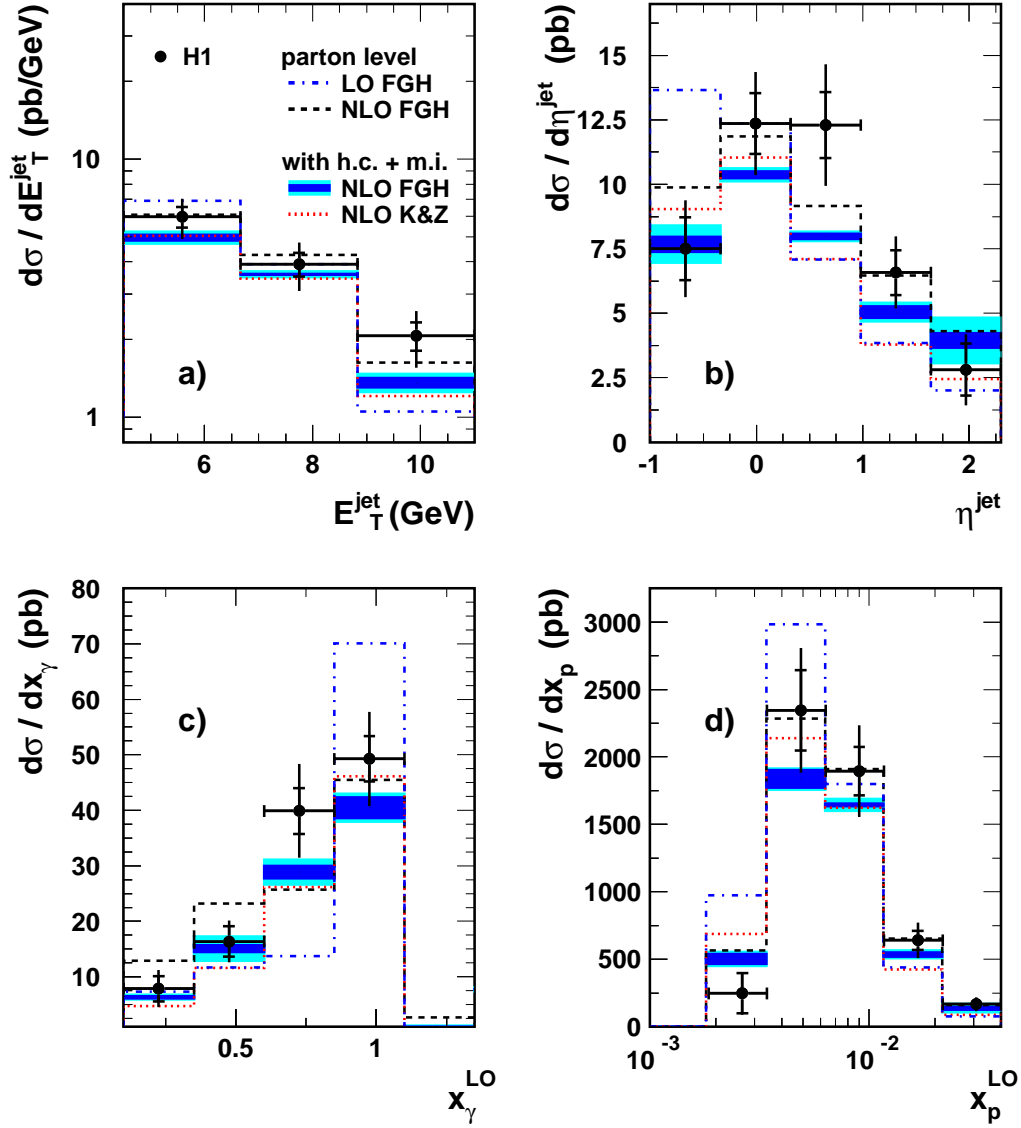}}
 \end{picture}
 \caption{
Prompt photon cross sections with an additional jet requirement
  ($E_T^{jet} > 4.5$ GeV, $-1 < \eta^{jet} < 2.3$)
 differential in
 $E_T^{jet}$, $\eta^{jet}$,
 $x_{\gamma}^{LO}$ and $x_p^{LO}$.
The data are compared with NLO pQCD calculations
 (K\&Z~\cite{Zembrzuski:2003nu}, dotted line and  
FGH~\cite{Fontannaz:2001ek,Fontannaz:2001nq} solid line).
The NLO results are  
corrected for hadronisation and 
multiple interaction (h.c.+ m.i.) effects (see text).
 For the error bands see Fig.~\ref{fig:res}.
The FGH results are also shown
 without corrections for h.c. and m.i. at NLO (dashed) and LO (dashed-dotted).
}
 \label {NLOscale}
\end{center}
\end{figure}

 Cross sections for the production of
 a prompt photon associated with a jet are presented
 in Figs.~\ref{fig:res}c,d
 as a function of the variables
    $E_T^{\gamma}$ and $\eta^{\gamma}$
 and in Fig.~\ref{NLOscale} as function of
 $E_T^{jet}, \eta^{jet},
 x_{\gamma}^{LO}$ and  $x_{p}^{LO}$.
 The estimators
 $x_{\gamma}^{LO} = E_T^{\gamma}(e^{-\eta^{jet}} + e^{-\eta^{\gamma}})/2yE_e$
 and $x_{p}^{LO} = E_T^{\gamma}(e^{\eta^{jet}} + e^{\eta^{\gamma}})/2E_p$
 are taken
  for the momentum fractions of constituents of the incident photon
 and proton,
  respectively, participating in the hard process.
 These observables make explicit use only of the
 photon energy, which is better measured than the jet energy.  
 They are thus most easily interpreted in 
 the leading order (LO)
 approximation
where the outgoing partons from the hard interaction,
 and correspondingly the reconstructed photon and jet,
 have equal transverse momenta. 
  The use of the variable  $x_{\gamma}^{LO}$
 was recommended 
 in~\cite{Aurenche:1994in,Fontannaz:2001nq}
 to reduce infrared sensitivity.
 The variable  $x_{p}^{LO}$ is discussed e.g. in~\cite{Fontannaz:2002nu}.
     
The data are compared
 with the pQCD calculations.
 In contrast to the inclusive case, both NLO calculations are
 consistent with the data in most bins.
 The NLO/LO correction ratios for \dsdeta
 are more moderate than in the inclusive case,
 ranging from about 0.95 to 1.25 with increasing \etag for FGH
   (Fig.~\ref{fig:res}d),
 but are still large in some bins of other distributions,
as shown in  Fig.~\ref{NLOscale}.
 The hadronic and m.i. corrections, which are applied for
 both NLO calculations,
 improve the description of the data only in some regions,
 such as $\eta^{\gamma} > 0.6$,
 $\eta^{jet} < -0.3$ and $x_{\gamma}^{LO} < 0.6$.
 The $x_{\gamma}^{LO}$ distribution  (Fig.~\ref{NLOscale}c)
 is particularly sensitive to the 
 photon structure function.
Using the  GRV parameterisation~\cite{Gluck:1991jc,Gluck:1991ee},
 the K\&Z prediction increases by about $4\%$ for  $x_{\gamma}^{LO} > 0.85$
 and by about 20\% at $x_{\gamma}^{LO} < 0.85$ where resolved photon 
 contributions dominate, leading to a somewhat improved description
 of the data.   

 Further understanding of the dynamics of the process and in particular
 of the effect of higher order gluon emissions beyond the diagrams
 shown in Fig.~\ref{feyn}  
 may be obtained
 from the transverse correlation between
 the prompt photon and the jet.
 The distribution of the component of the prompt photon's momentum
 perpendicular
  to the jet direction in the transverse plane,  
  $$p_{\bot} \equiv \; \mid\vec p_T^{\;\gamma} \times \vec p_T^{\;jet}\mid /
\mid\vec p_T^{\;jet}\mid \;= E_T^{\gamma} \cdot \sin(\Delta \phi)\;,$$
 where $\Delta \phi$ is the difference in azimuth between
 the photon and the jet,
 is determined by 
 higher order effects as $p_{\bot}$ is zero at leading order, where 
 the prompt photon and the jet are back-to-back in the 
 transverse plane. 

\begin{figure}[ht] \unitlength 1cm
\begin{center}
\begin{picture}(16,19.)
 \put(0.5,9.){\includegraphics[width=0.9\textwidth]{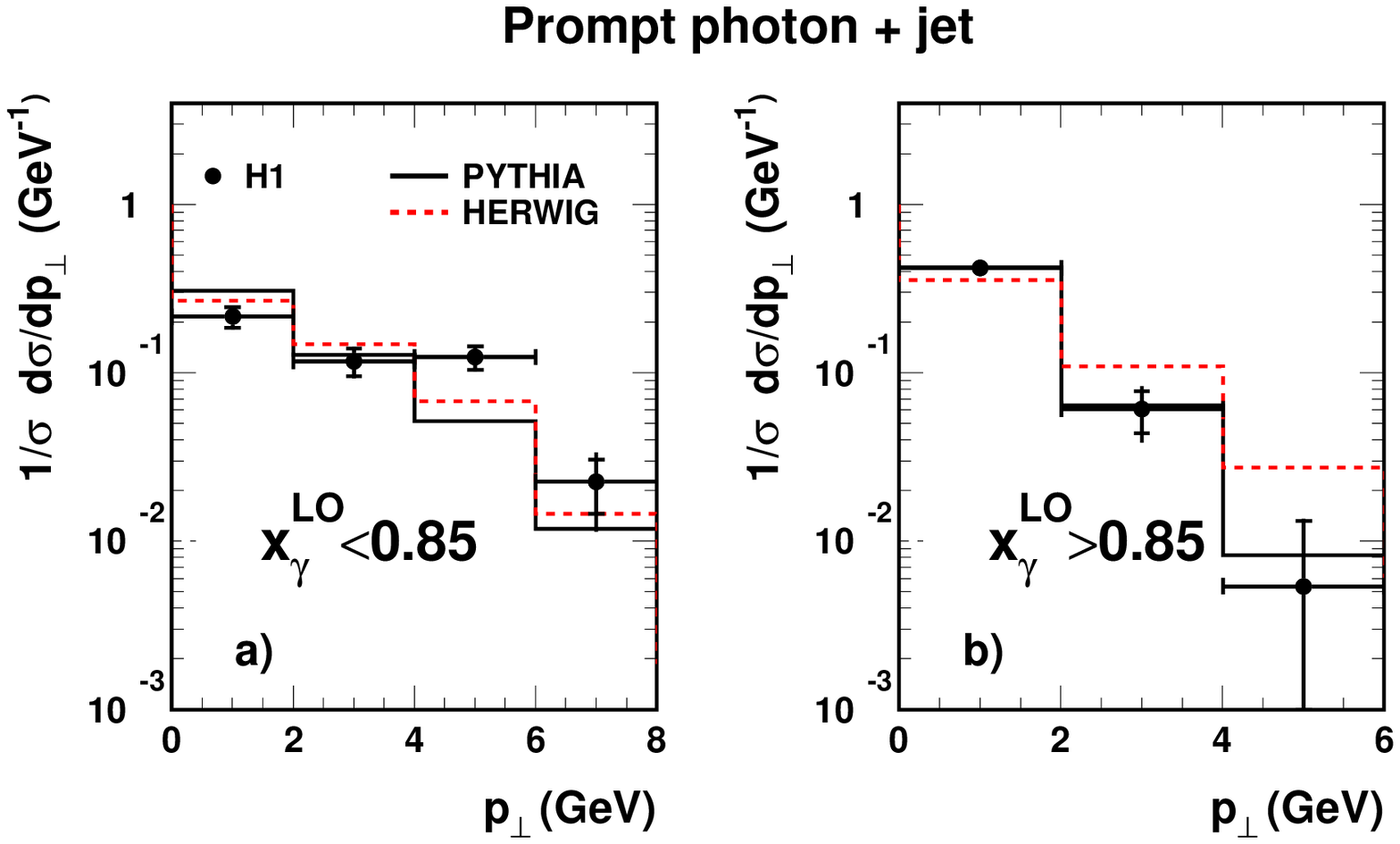}}
 \put(0.5,0.){\includegraphics[width=0.9\textwidth]{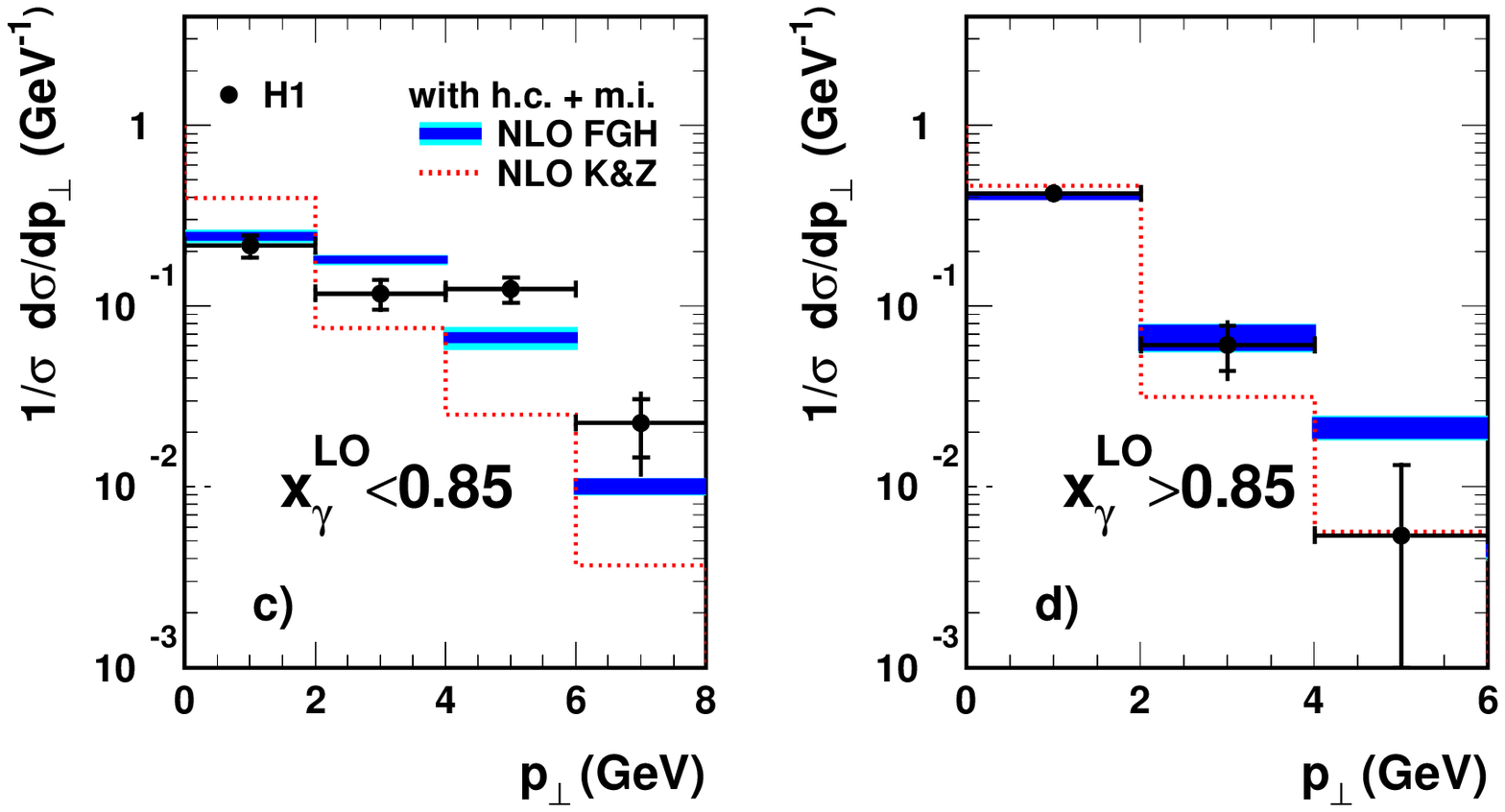}}
 \end{picture}
 \caption{
Distribution of the prompt photon momentum component,
perpendicular to the jet direction in the transverse plane, for 
$x^{LO}_{\gamma} < 0.85$, a) and c), and $x^{LO}_{\gamma} > 0.85$, b) and d).
In a) and b) the data are compared with PYTHIA (solid line)
 and HERWIG (dashed line).  
In c) and d) the data 
 are compared with NLO pQCD calculations  
(K\&Z~\cite{Zembrzuski:2003nu}, dotted line, and
FGH~\cite{Fontannaz:2001ek,Fontannaz:2001nq}, solid line).
The NLO results are 
 corrected for hadronisation and 
  multiple interactions.
For the error bands see Fig.~\ref{fig:res}. 
}  
 \label {pperp}
\end{center}
\end{figure}

 The
 normalised
  $p_{\bot}$ distribution is shown in Fig.~\ref{pperp} separately
  for the regions $x^{LO}_{\gamma} > 0.85$ and $x^{LO}_{\gamma} < 0.85$,
  where direct and resolved photon induced
  processes dominate, respectively
(contributing about 80\% in each case according to the FGH
calculation). The FGH NLO prediction gives a better description
of the $p_{\bot}$ distributions than K\&Z. This is due, at least
in part, to the differences between the QCD corrections for
resolved photon interactions in the two calculations. For
$x^{LO}_{\gamma} > 0.85$, the data are quite well described
by PYTHIA, whereas the HERWIG prediction is somewhat harder than
that of the data. For $x^{LO}_{\gamma} < 0.85$, the measured
distribution is reasonably well described by both Monte Carlo
models, except in the region of $p_{\bot} \approx 5$~GeV.
The $p_{\bot}$ distribution at large $x^{\gamma}$ has previously
been used \cite{Chekanov:2001aq} to extract information on an
intrinsic $k_t$ of the initial state partons of the proton.
 However, the large differences
between the predictions of the various NLO calculations and Monte
Carlo models in the present comparisons do not allow a reliable
statement to be made.

\section{Conclusions}

The  photoproduction of prompt photons,
    both inclusively and  
    associated with jets, is studied.
 The measured $\eta^{\gamma}$   and $E_T^{\gamma}$ distributions of
 the inclusive prompt photons are reasonably well described in shape 
 by NLO pQCD calculations,
 but after corrections for hadronisation and multiple interactions 
 the predictions are
$30\% - 40\%$ 
 below the data.
  The cross sections predicted by the  
  PYTHIA and HERWIG event generators describe the data distributions
 well in shape
  with normalisations that are low by about $40\% - 50\%$.

  For prompt photons associated with a jet, the data
 are somewhat better described
 by the NLO calculations including corrections for hadronisation
 and multiple interactions. 
 This, together with the fact that the NLO corrections are
   smaller on average
  than in the inclusive case,
  suggests that contributions beyond NLO are less important
 if an energetic jet is selected together with
  the prompt photon. 

 The distribution of  $p_{\bot}$,
 the component of the prompt photon's momentum
 perpendicular
  to the jet direction in the transverse plane, is sensitive
 to effects beyond LO.
  The PYTHIA generator describes the normalised
  $p_{\bot}$ distributions quite well,
 whereas HERWIG predicts too hard a 
   $p_{\bot}$ distribution at large $x_{\gamma}^{LO}$, where direct photon
  interactions dominate.
Particularly at low  $x_{\gamma}^{LO}$, the  $p_{\bot}$ distribution
is better described by the  NLO calculations
  if  NLO QCD corrections are also applied 
  in the case of the resolved photon interactions.

\vspace*{14.25pt}

\section*{Acknowledgements}

We are grateful to the HERA machine group whose outstanding
efforts have made this experiment possible. 
We thank
the engineers and technicians for their work in constructing and
maintaining the H1 detector, our funding agencies for 
financial support, the
DESY technical staff for continual assistance
and the DESY directorate for support and for the
hospitality which they extend to the non DESY 
members of the collaboration.
 We are grateful to Gudrun Heinrich and Andrzej Zembrzuski
 for discussions and help with 
 the NLO QCD calculations.

\vspace*{14.25pt} 

%
%


\begin{table}[htb]
\begin{center}
\begin{tabular}{|c|c|c|}
\hline
  \Etg (GeV)  & \dsdE (pb/GeV) &h.c.+ m.i\\ 
 \hline
 5.0 $-$        5.8 &       17.0 $\pm       1.8 \pm$       2.9  & 0.75 \\
 5.8 $-$        6.7 &       11.6 $\pm       1.2 \pm$       1.9 &  0.78 \\
 6.7 $-$        7.5 &    8.3 $\pm       0.9 \pm$       1.3 & 0.84 \\
 7.5 $-$        8.3 &    6.67 $\pm       0.69 \pm$       0.85 &  0.88 \\
 8.3 $-$        9.2 &    5.46 $\pm       0.55 \pm$       0.93 &  0.91 \\
 9.2 $-$       10.0 &    2.42 $\pm       0.48 \pm$       0.57 &  0.91 \\
\hline
\hline
   \etag   & \dsdeta (pb)  &h.c.+ m.i. \\ 

 \hline
 -1.0 $-$      -0.7 &       26.8 $\pm       2.6 \pm$       4.2  &  0.94 \\
 -0.7 $-$        -0.4 &       29.1 $\pm       3.4 \pm$       5.2 & 0.87 \\
 -0.4 $-$        0.0 &       26.9 $\pm       3.0 \pm$       4.3 &  0.84 \\
  0.0 $-$        0.3 &       24.7 $\pm       2.7 \pm$       3.4 &  0.75 \\
  0.3 $-$        0.6 &       21.2 $\pm       1.9 \pm$       3.5 &  0.71 \\
  0.6 $-$        0.9 &       11.0 $\pm       1.6 \pm$       2.2 &  0.65 \\
\hline
 \end{tabular}
\caption{ Inclusive prompt photon differential cross sections
\dsdE for  $-1 < \eta^{\gamma} < 0.9$ 
and \dsdeta for $5 <$ \Etg $< 10$ GeV 
 with $\sqrt{s} = 319~\GeV$ and $0.2 < y < 0.7$.
   The first error is statistical, the second systematic.
 The correction factors applied to the NLO calculations for hadronisation
 and multiple interactions (h.c. + m.i.) are also given. 
}
\label{metab1}
\end{center}
\end{table}

\clearpage

\begin{table}[htb]
\begin{center}
\begin{tabular}{|c|c|c|}
\hline
  \Etg (GeV)  & \dsdE (pb/GeV) &h.c.+ m.i \\ 
 \hline
 5.0 $-$        5.8 &        8.9 $\pm    1.2 \pm$    2.4 & 0.77 \\
 5.8 $-$        6.7 &        6.24 $\pm    0.90 \pm$    1.02 & 0.80 \\
 6.7 $-$        7.5 &        6.10 $\pm    0.77 \pm$    0.97 & 0.85 \\
 7.5 $-$        8.3 &        5.28 $\pm    0.64 \pm$       0.71 & 0.89 \\
 8.3 $-$        9.2 &        3.98 $\pm    0.51 \pm$       0.71 & 0.93 \\
 9.2 $-$       10.0 &        2.30 $\pm    0.48 \pm$       0.52 & 0.91 \\
\hline
\hline
   \etag   & \dsdeta (pb) &h.c.+ m.i. \\ 
 \hline
 -1.0 $-$       -0.7 &       16.2 $\pm       2.1 \pm$       2.8 & 0.95\\
 -0.7 $-$       -0.4 &       18.2 $\pm       2.5 \pm$       3.2 & 0.89\\
 -0.4 $-$        0.0 &       18.2 $\pm       2.4 \pm$       2.9 & 0.87\\
  0.0 $-$        0.3 &       13.7 $\pm       2.0 \pm$       2.2 & 0.77\\
  0.3 $-$        0.6 &       14.5 $\pm       1.5 \pm$       2.6 & 0.76\\
  0.6 $-$        0.9 &        6.8 $\pm       1.3 \pm$       1.6 & 0.71\\
\hline
\hline
 $E_T^{jet}$ (GeV) & ${\rm d}\sigma / {\rm d} E_T^{jet}$ (pb/GeV) &h.c.+ m.i.\\ 
\hline
 4.5 $-$        6.7 &        5.97 $\pm       0.54 \pm$       0.91 & 0.82\\
 6.7 $-$        8.8 &        3.92 $\pm       0.42 \pm$       0.72 & 0.84\\
 8.8 $-$       11.0 &        2.07 $\pm       0.26 \pm$       0.44 & 0.83\\
\hline
 \hline
  $ \eta^{jet}$   & ${\rm d}\sigma / {\rm d} \eta^{jet}$ (pb) &h.c.+ m.i. \\
\hline 
 -1.0 $-$       -0.3 &        7.5 $\pm       1.2 \pm$       1.4 & 0.77\\
 -0.3 $-$        0.3 &       12.4 $\pm       1.2 \pm$       1.6 & 0.88\\
 0.3 $-$        1.0 &       12.3 $\pm       1.3 \pm$       2.0 & 0.86\\
 1.0 $-$        1.6 &        6.58 $\pm       0.87 \pm$       1.10 & 0.78\\
 1.6 $-$        2.3 &        2.81 $\pm       1.01 \pm$       0.94 & 0.91\\
\hline
\hline
 $x_{\gamma}^{LO}$ & ${\rm d}\sigma / {\rm d} x_{\gamma}^{LO}$ (pb)  &h.c.+ m.i.\\
\hline
  0.1 $-$        0.3 &        7.8 $\pm       2.2 \pm$       2.5 & 0.50\\
  0.3 $-$        0.6 &       16.4 $\pm       2.8 \pm$       2.6 & 0.65\\
  0.6 $-$        0.9 &       39.9 $\pm       4.1 \pm$       7.4 & 1.1 \\
  0.9 $-$        1.1 &       49.3 $\pm       4.1 \pm$       7.4 & 0.88\\
\hline
\hline
 $x_p^{LO}$ & ${\rm d}\sigma / {\rm d} x_p^{LO}$ (pb)
&h.c.+ m.i. \\
 \hline
  0.0018 $-$   0.0034 &       247 $\pm       149 \pm$        51 & 0.84\\ 
  0.0034 $-$   0.0063 &      2350 $\pm       300 \pm$       350 & 0.80\\
  0.0063 $-$   0.0120  &      1900 $\pm       180 \pm$       290 & 0.86\\
  0.0120 $-$   0.0220 &       640 $\pm        70 \pm$       113 & 0.81\\
  0.0220 $-$   0.0400 &       167 $\pm        31 \pm$        39 & 0.84 \\
\hline
 \end{tabular}
\caption{
Prompt photon cross sections with an additional jet requirement
  ($E_T^{jet} > 4.5$ GeV, $-1 < \eta^{jet} < 2.3$)
   differential in
$E_T^{\gamma}$, $\eta^{\gamma}$, $E_T^{jet}$, $\eta^{jet}$,
 $x_{\gamma}^{LO}$ and $x_p^{LO}$
for  $-1 < \eta^{\gamma} < 0.9$ 
and $5 <$ \Etg $< 10$ GeV 
 with $\sqrt{s} = 319~\GeV$ and $0.2 < y < 0.7$.
The first error is statistical, the second systematic.
 The correction factors applied to the NLO calculations for hadronisation
 and multiple interactions (h.c. + m.i.) are also given.  
}
\label{metab2}
\end{center}
\end{table}

\begin{table}[htb]
\begin{center}
\begin{tabular}{|c|c|}
\hline
  $p_{\perp}$ (GeV)  & $1/\sigma \;{\rm d}\sigma / {\rm d}p_{\perp}
 ($GeV$^{-1}),\;
  x_{\gamma}^{LO} < 0.85$ \\ 
\hline
        0 $-$         2 &      0.216 $\pm      0.030 \pm$      0.015 \\
        2 $-$         4 &      0.117 $\pm      0.022 \pm$      0.011 \\
        4 $-$         6 &      0.124 $\pm      0.019 \pm$      0.011 \\
        6 $-$         8 &      0.0225 $\pm      0.0081 \pm$      0.0077 \\
\hline
\hline
  $p_{\perp}$ (GeV)  & $1/\sigma \;{\rm d}\sigma / {\rm d}p_{\perp}
 ($GeV$^{-1}),\;
 x_{\gamma}^{LO} > 0.85$ \\

 \hline
        0 $-$         2 &      0.420 $\pm      0.033 \pm$      0.024 \\
        2 $-$         4 &      0.061 $\pm      0.017 \pm$      0.014 \\
        4 $-$         6 &      0.0054 $\pm      0.0078 \pm$      0.0026 \\
\hline
 \end{tabular}
\caption{
Normalised cross sections differential in the prompt photon momentum
 component
perpendicular to the jet direction in the transverse plane, for  
$x^{LO}_{\gamma} < 0.85$ and $x^{LO}_{\gamma} > 0.85$ 
with $5 <$ \Etg $< 10$ GeV,
$-1 < \eta^{\gamma} < 0.9$,
  $E_T^{jet} > 4.5$ GeV and $-1 < \eta^{jet} < 2.3$
 with $\sqrt{s} = 319~\GeV$ and $0.2 < y < 0.7$.
The first error is statistical, the second systematic. 
}
\label{metab3}
\end{center}
\end{table}


\clearpage

\end{document}